\documentclass[letterpaper,10pt,conference]{ieeeconf}

\usepackage[noadjust]{cite}
\IEEEoverridecommandlockouts                              
\usepackage[pdftex]{graphicx}
\DeclareGraphicsExtensions{.pdf}
\usepackage{amsmath} 
\usepackage{amssymb}  
\usepackage{booktabs}
\usepackage{upgreek}
\usepackage{footnote}
\makesavenoteenv{tabular}
\makesavenoteenv{table}
\usepackage[linesnumbered,algoruled,vlined]{algorithm2e}
\usepackage{url}
\usepackage{balance}
\usepackage{caption, booktabs}
\usepackage[flushleft]{threeparttable}

\newcommand{\ee}{\textrm{e}}
\newcommand{\jj}{\textrm{j}}
\newcommand{\ud}{\textrm{d}}

\title{\LARGE \bf
  Machine learning without a feature set for detecting bursts in the EEG of preterm infants 
}

\author{John M. O'~Toole and Geraldine B.~Boylan
\thanks{Supported by Science Foundation Ireland
(15/SIRG/3580 and INFANT-12/RC/2272).}%
\thanks{Authors are with the Irish Centre for Fetal and Neonatal Translational
   Research (INFANT), University College Cork, Ireland. (email: {\small \tt jotoole@ucc.ie})}%
}

\begin{document}

\maketitle
\thispagestyle{empty}
\pagestyle{empty}

\begin{abstract}
  Deep neural networks enable learning directly on the data without the domain knowledge
  needed to construct a feature set.
  This approach has been extremely successful in almost all machine learning applications.
  We propose a new framework that also learns directly from the data, without extracting a
  feature set.
  We apply this framework to detecting bursts in the EEG of premature infants. 
  The EEG is recorded within days of birth in a cohort of infants without significant
  brain injury and born $<$30 weeks of gestation.
  The method first transforms the time-domain signal to the time--frequency domain and
  then trains a machine learning method, a gradient boosting machine, on each time-slice
  of the time--frequency distribution.
  We control for oversampling the time--frequency distribution with a significant
  reduction ($<$1\%) in memory and computational complexity.
  The proposed method achieves similar accuracy to an existing multi-feature approach:
  area under the characteristic curve of 0.98 (with 95\% confidence interval of 0.96 to
  0.99), with a median sensitivity of 95\% and median specificity of 94\%.
  The proposed framework presents an accurate, simple, and computational efficient
  implementation as an alternative to both the deep learning approach and to the manual
  generation of a feature set.
\end{abstract}

\section{Introduction}

Deep learning has created a paradigm shift in machine learning classification
\cite{Lecun2015,Miotto2017,Min2017}.
The key advantage of the deep learning framework is that specific domain-knowledge is not
needed to engineer features from the raw data. 
Better still, classification performance has outstripped other machine-learning approaches
in many diverse applications \cite{Lecun2015}. 
A deep neural network is trained directly on the data, such as an image or time-domain
signal, and inherently extracts features in layers of abstraction through the
network. This enables end-to-end training of the both the feature extraction and
classification process, with remarkable improvements in accuracy
\cite{Lecun2015,Miotto2017,Min2017}.

Deep neural networks may not suit every application, however. To start, they typically
require extra computational resources (graphical processing units) to train a network
within a reasonable time-frame. 
Second, designing the architecture of a deep neural network is a complicated process, as
no there is no one-size-fits all model. This process can be as elaborate as the often
derided \emph{hand-crafted features} approach: there may be as much artisan effort in
designing a network as designing a feature set. Third, because of the end-to-end
optimisation of feature extraction and classification, the networks add an extra layer of
obfuscation making if difficult to know what characteristics of the data are important in
the classification task. 
This can present a problem in some fields, such as medicine \cite{Miotto2017}. Fourth,
deep learning has excelled in problems with big data sets. 
Our area of study, brain monitoring in critical-ill infants, uses valuable but small-size
data sets.
We have not yet witnessed significant improvements with deep learning in this area
compared to well constructed feature sets; recent methods in seizure detection, for
example, contrast this difference \cite{Ansari2018a, Tapani2018}.

We present a new framework that aims to address the limitations of deep learning by
challenging the idea that only deep neural networks can learn directly from the data,
without the need for feature extraction. This new framework firsts transforms the signal
to the time--frequency domain and then applies a robust machine learning method directly
on the instantaneous frequency representation. The advantage of this approach is that the
signal is not dependent on time and therefore affords a level of abstraction in the
frequency domain. This abstraction is controlled by parameters in the smoothing kernel of
the quadratic time--frequency distribution. We apply this new framework to detecting the
bursts and inter-bursts in the EEG of premature infants \cite{OToole2017}. The length of
inter-bursts is an important marker of brain maturity \cite{Pavlidis2017b} and is used in
segmenting the EEG for automated applications \cite{OToole2016b}.

\section{Methods}
\label{sec:methods}

\subsection{EEG recordings}
\label{sec:eeg-recordings}

This study uses EEG data acquired, within days of birth, from 36 infants born $<$30 weeks
of gestation.
Informed and written consent was obtained before EEG recording. 
Approval for the study was obtained from the Cork Research Ethics Committee of Cork
Teaching Hospitals, Ireland.  
This is the same data set presented in \cite{OToole2017}.
EEG was recorded using a modified 10--20 system of electrode configuration covering the
frontal, central, temporal, and occipital regions.  
Ten minute epochs, with little artefact, were pruned from the long-duration EEG. 
Two EEG experts independently annotated 1 channel of bursts and inter-bursts for the 10
minute epochs.
Burst were defined as any EEG activity not labelled as inter-bursts. 
Thus continuous activity was also labelled as bursts \cite{OToole2017}.
For training and testing, we use the consensus annotation between the 2 reviewers.

The EEG was downsampled from 256~Hz to 64~Hz after applying an anti-aliasing filtering. 
The signal was then band-passed filtered within 0.5 to 30~Hz using a infinite impulse
response (IIR) filter. 
The filter was implemented from a 5th-order Butterworth design in a forward--backwards
procedure to enforce a zero-phase response.

\subsection{From time to time--frequency}
\label{sec:from-time-time}

We use a quadratic time--frequency distribution (TFD) to transfer the EEG signal $x(t)$ to
time--frequency domain $\rho(t,f)$. 
We first segment $x(t)$ into short-duration epochs of 32 seconds and then generate a
separable-kernel TFD. 
This TFD has the form
\begin{equation}
  \label{eq:1}
  \rho(t,f)=W(t,f) \underset{t}{*} \underset{f}{*}\; \gamma(t,f)
\end{equation}
where $W(t,f)$ is the Wigner--Ville distribution and $\gamma(t,f)$ is the kernel, a two-dimensional
smoothing function \cite{Boashash2013}.  The Wigner--Ville distribution is a function of $x(t)$
\begin{equation*}
  W(t,f)=\int_{-\infty}^{\infty} z(t+\tfrac{\tau}{2}) z^*(t-\tfrac{\tau}{2})%
  \ee^{-\jj2\uppi \tau f} \ud\tau
\end{equation*}
where $z(t)$ is analytic associate of signal $x(t)$ and ${z}^*(t)$ represents the complex
conjugate of $z(t)$ \cite{otoole2008}. The kernel, $\gamma(t,f)$ in \eqref{eq:1}, is a
two-dimensional smoothing function which suppresses the cross- and inner-terms from the
quadratic $W(t,f)$ function. 
For this application, we use a separable kernel of the form $\gamma(t,f) = G(t)H(f)$
\cite{Boashash2013}. This kernel is often defined as a low-pass filter in the Doppler--lag
domain $g(\nu)h(\tau)$, the two-dimensional Fourier transform of the time--frequency domain
\cite{Boashash2013}. 
We use a Hanning window for $h(\tau)$ and a Tukey window, with a parameter value 0.9, for
$g(\nu)$.

To enable efficient computation of the TFD in \eqref{eq:1}, we employ fast and
memory-efficient algorithms to avoid oversampling in the time and frequency direction
\cite{OToole2013,OToole2016}. 
The length for both the lag and Doppler windows is set to 61 samples. To avoid wrap-around
effects in time from the time--frequency convolution in \eqref{eq:1}, we extend the epoch
2 seconds in either direction. That is, we generate the TFD for a 36-second epoch.  
To avoid overlapping in the frequency direction, we interpolate the signal by a factor of
2 by zero-padding in the frequency domain.
After generating the TFD, 2-seconds at the start and end in the time-direction are
discarded from the TFD as are all frequencies from $f_s/4$ to $f_s/2$, where $f_s$ represents the
sampling frequency of the interpolated signal.
The next epoch is overlapped with the previous one so all EEG is analysed. 

\begin{figure*}
  \centering
  \includegraphics[scale=0.65]{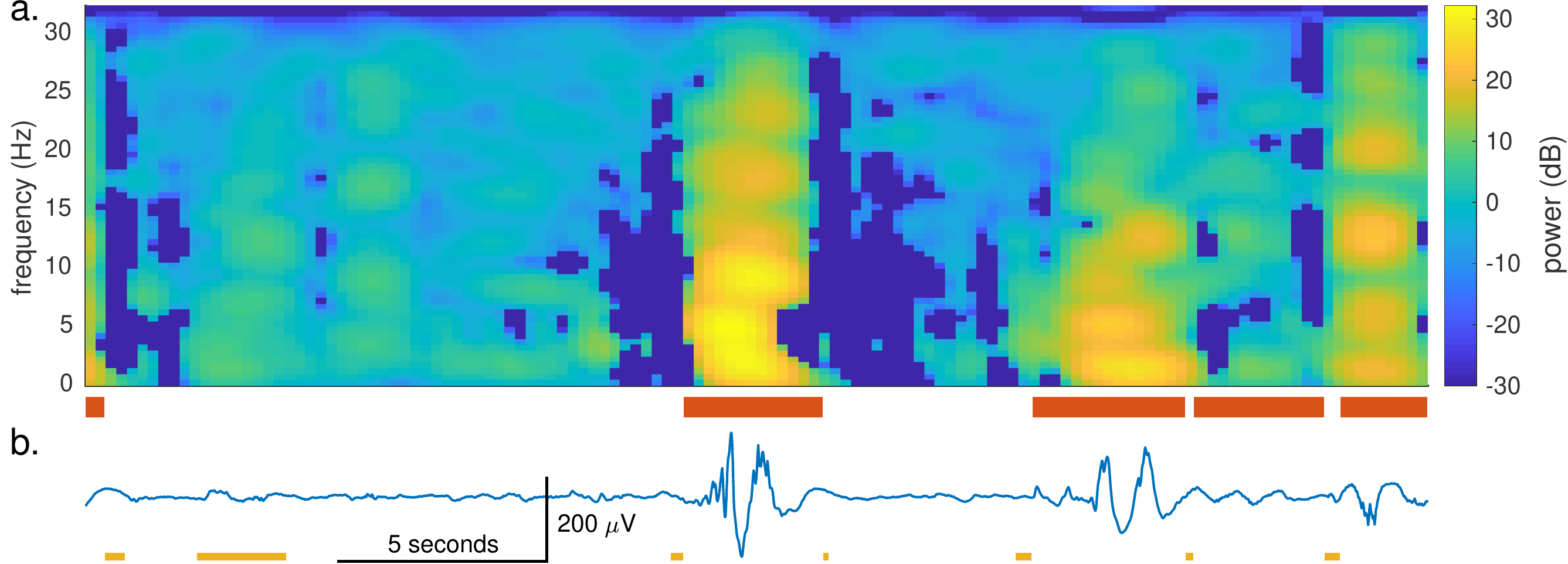} \vspace{0.2em}
  \caption{Time--frequency distribution (TFD) in (a) generated from EEG epoch in (b)
    containing bursts and inter-bursts. Thick red lines over the EEG represent the
    consensus annotation for the bursts.
    Thin yellow lines below the EEG represent periods where the 2 annotators did not
    agree. The TFD is a $128 \times 64$ matrix, much smaller than the full, but
    oversampled, $2,048\times 2,048$ TFD (the EEG epoch contains $2,048$ samples). The
    derivative of the signal is used to enhance high-frequency components in the TFD.}
  \label{fig:tfd_eg}
\end{figure*}

Before generating the TFD, a simple pre-whitening filter is used to emphasize higher
frequency components of the EEG signal which is otherwise dominated by high-amplitude
low-frequency components. 
This approach has worked well in other newborn EEG applications, for example see
\cite{Tapani2018}. 
We approximate the whitening filter by computing the derivative of signal using the
forward--finite difference method.
Fig.~\ref{fig:tfd_eg} shows an example TFD for an
epoch of EEG containing both bursts and inter-bursts.
Computer code to generate the efficient TFDs is available at
\url{https://github.com/otoolej/memeff_TFDs} (commit: 64daec4).  

To compare the computational performance of this efficient TFD implementation, we
calculate the number of arithmetical operations and memory required to compute the TFD. 
The separable-kernel TFD can be computed using
\begin{multline}
  \label{eq:2}
  C_{\textrm{eff}} = P_h (N\log_2 N + N_{\textrm{time}} \log_2 N_{\textrm{time}}) \\  
  + \frac{1}{2}N_{\textrm{time}} N_{\textrm{freq}} \log_2 N_{\textrm{freq}} 
\end{multline}
arithmetical operations (multiplications and additions) for the TFD of size (time $\times$
frequency) $N_{\textrm{time}}\times N_{\textrm{freq}}$ \cite{OToole2013}; $P_h$ is half
the length of the lag window $h(\tau)$. 
This calculation assumes that a fast-Fourier transform of a length-$N$ complex-valued
signal requires $N\log_2 N$ arithmetical operations. In comparison, the full (oversampled)
TFD of size $N \times N$ requires
\begin{equation}
  \label{eq:3}
  C_{\textrm{full}} = \frac{3N^2}{2} \log_2 N
\end{equation}
arithmetical operations.

\begin{figure}[ht]
  \centering
  \includegraphics[scale=0.77]{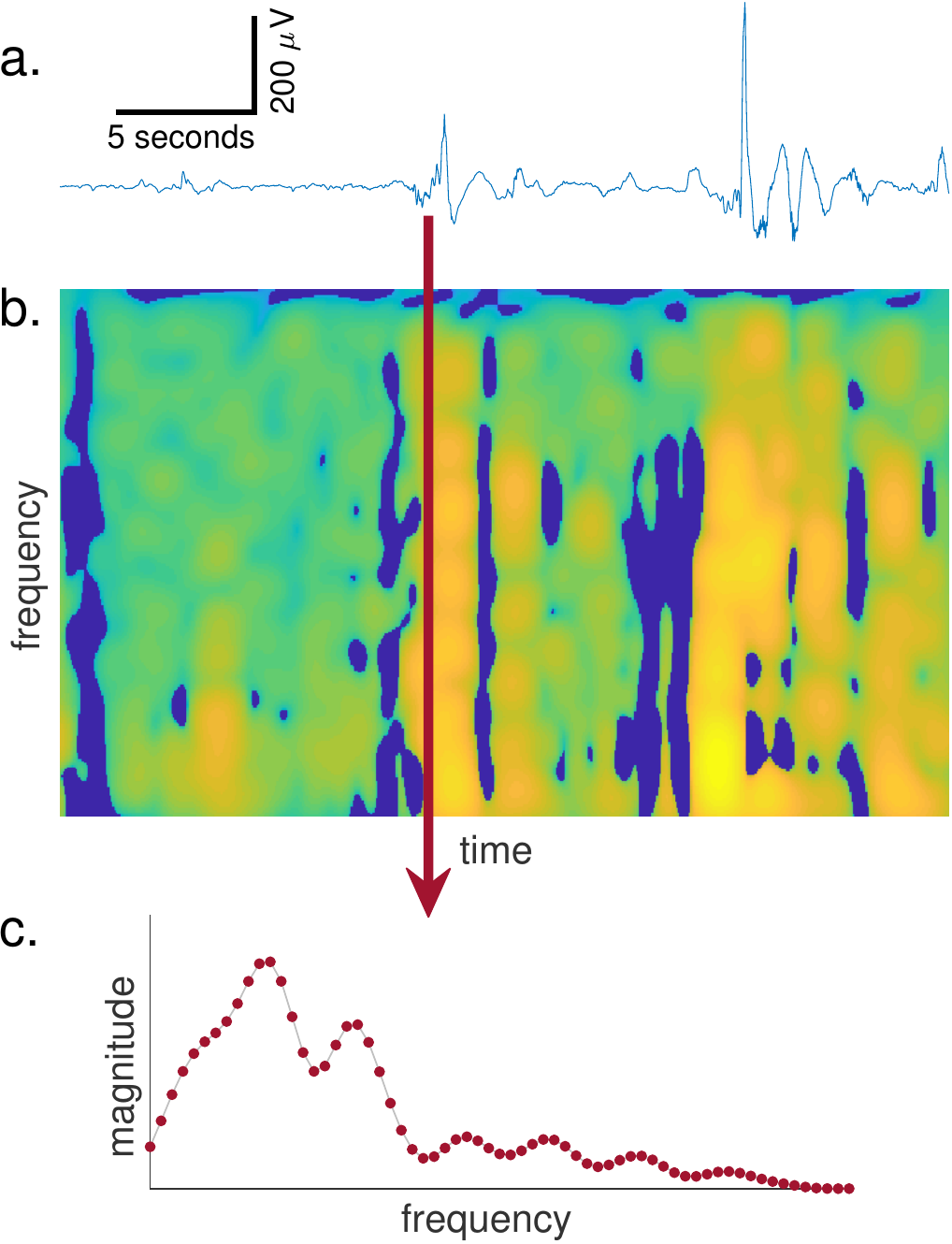}
  \caption{Time-slice in (c) of the time--frequency distribution (TFD) in (b), generated
    from the EEG epoch in (a). The red arrow illustrates the time location used to
    extract the time-slice from the TFD. 
    The 64 samples in (c) are the feature vector that represents 1/4 of a second of the EEG
    epoch.}
  \label{fig:process_eg}
\end{figure}

We set $N_{\textrm{time}} = 144$, which equates to a sampling frequency of 4~Hz, and
$N_{\textrm{freq}}=128$, which equates to frequency resolution of 1/2~Hz.  
This TFD is trimmed to a $128 \times 64$ matrix to represent the 32 second epoch
with frequencies $\leq 32$~Hz.
The $i$-th time-slice of the frequency-domain $\rho[n_i,k]$ representation is used as a
feature vector of $k=0,1,\ldots,63$ data points.  
Thus, for each 1/4 second of EEG, the feature vector $\rho[n_i,k]$ is used to train
and test the machine learning algorithm. 
Fig.~\ref{fig:process_eg} illustrates this process.  

\subsection{Machine learning}
\label{sec:shallow-learning}

We use a gradient boosting machine (GBM) with decision trees \cite{Friedman2001},
although the proposed framework should work with any robust classification method.  
The GBM is an ensemble technique which iteratively adds a sequence of weak learners, in
this case regression trees, to improve (boost) classification accuracy.  
Each new tree added tries to correct the mis-classifications of the previous tree, thus
iteratively improving performance.

The GBM was trained and tested on a leave-one-out cross-validation scheme. 
Data was segregated on each infant's EEG record for the cross validation.

The GBM was implemented using the \emph{XGBoost} algorithm \cite{Chen2016}. 
This computational efficient implementation of the GBM incorporates a regularised
objective function.  
We set $\gamma=10$, a regularisation parameter \cite{Chen2016}, to avoid over-fitting the
model.
We include regularisation because our data set contains a high level of variability
between infants and therefore between the training and testing data sets. 
This is because characteristics of the preterm EEG differs significantly with gestational
age \cite{Pavlidis2017b}, with our infants spanning from 23.4 to 29.7 weeks' gestational
age.  
Other parameters use default values: maximum number of trees was 100, a learning rate of
0.3, all of the data was considered when growing each tree, and the maximum depth of the
trees was 6.
We expect a class imbalance between the burst and inter-bursts of at least a factor of
2:1, with bursts defined here as including continuous activity; thus we set the parameter
to scale the class weights according to this ratio.  
Refining parameter selection, through nested cross-validation, will be implemented at a
future date.

Performance of the detector was assessed using the area under the receiver-operator
characteristic (AUC).  
Results are reported with 95\% confidence intervals (CI) generated using the jackknife
method.

Many burst detection methods are based on an energy measure \cite{Palmu2010b, Koolen2014,
  OToole2014}. 
To confirm that the proposed method is an improvement over this approach, we compute the
time-marginal of the TFD to generate a feature of instantaneous energy. 
We compare detection performance, using AUCs, for this feature compared to the performance
of the proposed machine-learning approach.
The comparison also demonstrates whether the frequency information of the TFD is important
as separate features rather than just using an energy (summation in this case) measure.  
We also compare the proposed method with an existing multi-feature support vector machine
approach \cite{OToole2017}. 
Comparisons report the median and 95\% CI difference in AUC with statistical significance
testing using the one-sided Mann--Whitney $U$-test.

\section{Results}
\label{sec:results}

Detection performance for the proposed method was on par with the multi-feature method:
median AUC of 0.982 for the proposed compared to 0.989 for the existing method---see
Table~\ref{tab:res_all}. 
Although the proposed approach does not significantly outperform the existing method
($p>0.999$), the median difference in AUC is $<$1\% and therefore below the limit for
engineering significance as defined elsewhere \cite{OToole2017}.
Setting a threshold of 0.5 for the GBM output gives a median (95\% CI) sensitivity of
94.6\% (81.3 to 99.6\%) with specificity of 93.8\% (64.1 to 99.2\%).

\begin{table}
  \centering
  \begin{threeparttable}
  \caption{Detection performance of the proposed method compared to both the existing
    multi-feature approach and the time-marginal of the time--frequency distribution.}
  \label{tab:res_all}  
  
    \begin{tabular}{llll}
    \toprule
    & AUC & \% difference & $p$-value$^{\dagger}$ \\
    & median (95\% CI) & median (95\% CI) \\
      \midrule
      TM-TFD & 0.933   &  5.20 (1.10 to 20.80)  & $<$0.001 \\
       & (0.772 to 0.973) &  & \\
      multi-feature & 0.989   & -0.78 (-2.10 to 0.23)  & $>$0.999 \\
        & (0.974 to 0.997) & & \\
      proposed      & 0.982   &                       &         \\
      & (0.960 to 0.991) &  & \\
    \bottomrule
  \end{tabular}
    \begin{tablenotes}
    \item {\footnotesize $^{\dagger}$one-sided Mann--Whitney $U$-test.  }
    \item {\footnotesize Key: AUC, area under the receiver operator characteristic; CI,
        confidence interval; TM-TFD, time-marginal of the time--frequency distribution}
     \end{tablenotes}
  \end{threeparttable}
\end{table}

To highlight the importance of the computational efficient implementation of the TFD,
Table~\ref{tab:tfd_comp} shows how controlling the oversampling of the full-TFD greatly
reduces both the required computations and memory. Computational load $C_{\textrm{eff}}$
is generated from \eqref{eq:2} for the $144\times 128$ efficient-TFD and
$C_{\textrm{full}}$ from \eqref{eq:3} for the $4,608 \times 4,608$ full-TFD. 
Here, $N=4,608$, $P_h=31$, $N_{\textrm{time}}=144$, and $N_{\textrm{freq}}=128$ for the
36-second epoch. On a 2019 desktop computer, the efficient-TFD for the 36 second epoch of
EEG is computed within 7 to 10 milliseconds requiring $<$0.5~MB of memory.

\begin{table}
  \centering
  \begin{threeparttable}
    \caption{Computational load and memory load for computing the time--frequency
      distribution (TFD) of a 36-second epoch sampled at 128~Hz. }
    \label{tab:tfd_comp}  
    \begin{tabular}{lrrr}
      \toprule
                                             & full-TFD & efficient-TFD  & \% reduction \\
      \midrule
      computations$^{\dagger}$ $\qquad\quad$ & 88,941,913    & 5,308,416 & 1.0          \\
      memory$^{\ddagger}$                    & 894,319       & 18,432    & 0.3          \\
      \bottomrule
    \end{tabular}
    \begin{tablenotes}
    \item {\footnotesize $^{\dagger}$ number of arithmetical operations (additions and
        multiplications). 
      }
    \item {\footnotesize $^{\ddagger}$ number of real-valued data points required to
        compute and store the TFD.}      
    \end{tablenotes}
  \end{threeparttable}
\end{table}

\section{Discussion and Conclusions}
\label{sec:conclusions}

We present a simple and computational efficient framework to use modern machine-learning
methods without the need to develop a feature set.
This is achieved by transforming the EEG signal to the time--frequency domain using
time-slices of the TFD as the feature vectors. 
These TFD time-slices are independent of time and therefore do not require that the
machine learning method learn all temporal mappings of the waveforms.  
Quadratic TFDs use all the signal information with a level of generalisation controlled by
the extent of the smoothing from the kernel. 
The more smoothing in time--frequency, the less cross- and inner-terms in the TFD
\cite{Boashash2013}. 
These cross- and inner-terms relate to relative-phase information between signal
components. 
Therefore a smoothed TFD may represent different signals but with the same components. 
By varying the kernel we can transition from a Wigner--Ville distribution---which retains
all of the signal information and can be inverted back to the time-domain signal up to a
constant phase factor---to the Spectrogram, which is a positive TFD without cross- or
inner-terms.

For our application here, we applied a separable kernel to generate a non-positive TFD,
meaning some cross- and inner-terms remain. 
The separable kernel has 2 parameters which were set to arbitrary values. 
A better approach may be to optimise these parameters within a nested cross-validation
scheme.  
In addition, the method could also benefit from tuning the hyper-parameters of the GBM. 
We leave both of these optimisation aspects to future work.

In conclusion, we present a method to detect bursts in the EEG of preterm infants. 
This framework uses a common machine learning method, the GBM, without the need to develop
a specific feature set. 
We overcome the computational burden associated with quadratic TFDs by using efficient
algorithms that require only 1\% of the memory and computational load required by standard
algorithms. 
The proposed method has similar high-performance to a multi-feature machine-learning
method developed from specific domain knowledge \cite{OToole2017}. 
Detecting bursts in the EEG of preterm infants is an important stage in developing fully
automated brain monitoring technologies for this vulnerable population.


\end{document}